\documentclass[preprint2,10.5pt]{aastex}
\begin{document}

\title{\large{\rm{CONCERNING THE CLASSICAL CEPHEID {\it VI}$_C$ WESENHEIT FUNCTION'S STRONG METALLICITY DEPENDENCE}}}
\author{D. Majaess$^1$, D. Turner$^1$, W. Gieren$^2$}
\affil{$^1$ Saint Mary's University, Halifax, Nova Scotia, Canada}
\affil{$^2$ Universidad de Concepci\'on, Concepci\'on, Chile.}
\email{dmajaess@cygnus.smu.ca}

\begin{abstract}
Evidence is presented which supports findings that the classical Cepheid {\it VI}$_C$ period-Wesenheit function is relatively insensitive to metallicity. The viability of a recently advocated strong metallicity dependence was evaluated by applying the proposed correction ($\gamma=-0.8$ mag/dex) to distances established for the Magellanic Clouds via a Galactic {\it VI}$_C$ Wesenheit calibration, which is anchored to ten nearby classical Cepheids with measured HST parallaxes.  The resulting $\gamma$-corrected distances for the Magellanic Clouds (e.g., SMC, $\mu_{0,\gamma}\sim18.3$) are in significant disagreement with that established from a mean of $>300$ published estimates (NED-D), and a universal Wesenheit template featuring eleven $\delta$ Scuti, SX Phe, RR Lyrae, and Type II Cepheid variables with HST/Hipparcos parallaxes.  Conversely, adopting a null correction (i.e., $\gamma=0$ mag/dex) consolidates the estimates.  In tandem with existing evidence, the results imply that variations in chemical composition among Cepheids are a comparatively negligible source of uncertainty for $W_{VI_c}$-based extragalactic distances and determinations of $H_0$.  A new approach is described which aims to provide additional Galactic Cepheid calibrators to facilitate subsequent assessments of the {\it VI}$_C$ Wesenheit function's relative (in)sensitivity to abundance changes.  VVV/UKIDSS/2MASS $JHK_s$ photometry for clusters in spiral arms shall be employed to establish a precise galactic longitude-distance relation, which can be applied in certain cases to determine the absolute Wesenheit magnitudes for younger Cepheids.
\end{abstract}


\section{{\rm \footnotesize INTRODUCTION}}
Classical Cepheids are integral to the establishment of the Galactic and extragalactic distance scales \citep{pg04,tu10}, and the selection of a cosmological model \citep{mr09,fm10}. Consequently, it is imperative to assess the effect of metallicity on {\it VI}$_C$ Wesenheit classical Cepheid relations.  In particular, are abundance differences between Cepheids comprising the calibration and target population important? Certain researchers advocate that a sizable correction is necessary when establishing the distance to benchmark metal-poor classical Cepheids in the Magellanic Clouds via a {\it VI}$_C$ Wesenheit calibration tied to solar-abundance Galactic Cepheids.  The dependence of the {\it VI}$_C$ Wesenheit function on chemical composition is typically assessed by: 
\begin{enumerate}
\item Evaluating the Wesenheit slopes inferred from classical Cepheids in galaxies spanning a sizable abundance baseline.  A pertinent example being to examine solar to metal-poor classical Cepheids in the Milky Way, LMC, NGC 6822, SMC, and IC 1613.  The galaxies are listed in order of decreasing metal abundance, and span $\Delta \rm{[Fe/H]}\sim1$ \citep{lu98,ud01,mot06,ta07}. 
\item Comparing whether the {\it VI}$_C$ Wesenheit magnitude offset between differing classes of pulsating stars is insensitive to the galaxy sampled.  That may be evaluated by examining differences in Wesenheit space between RR Lyrae variables, $\delta$ Scuti variables (SX Phe variables), Type II Cepheids, and classical Cepheids in the Galaxy, LMC, SMC, and IC 1613.  Equivalent offsets in the absence of metallicity corrections imply that {\it VI}$_C$ Wesenheit functions are insensitive to abundance changes.  Similarly, comparing the mean color-excess inferred from various standard candles at a common zero-point (e.g., IC 1613) likewise enables a determination of the impact of metallicity on that parameter, although marginal differences may arise since classical Cepheids (population I objects) are often located in dustier regions. Furthermore, extinction estimates inferred from period-reddening ($VI_c$) relations can be compared to DIRBE/IRAS dust maps to constrain the metallicity dependence.      
\item Exploiting the galactocentric metallicity gradient to deduce the Wesenheit magnitude offset between classical Cepheids observed in the outer (metal-poor) and central (metal-rich) regions of a particular galaxy.  However, a degeneracy emerges which complicates the analysis since the surface brightness and stellar density increase toward the central (metal-rich) regions, and thus photometric contamination (blending/crowding) becomes significant.  Indeed, it is argued here that the (spurious) brightening ($W_{VI_c}$) of extragalactic Cepheids as a function of decreasing galactocentric distance \citep{ke98} is direct empirical evidence of photometric contamination.
\item Published metallicity corrections are evaluated by applying them to distances established for the Magellanic Clouds using a Galactic classical Cepheid calibration \citep[e.g.,][]{ma09b}.  The aim is to assess whether the metallicity corrected distances match expectations for the Magellanic Clouds as established from $\ge 3\times 10^2$ published estimates (e.g., SN1987A, eclipsing binaries, RR Lyrae variables, etc.).  
\end{enumerate}
In this study, evaluation (4) is conducted using the sizable metallicity effect\footnote{\citet[][submitted]{ge11} likewise favour a sizable metallicity dependence.  The reader is referred to their comprehensive survey.} ($\gamma\sim-0.8$ mag/dex) proposed by \citet{ss10}.  \citet{ss10} inferred that estimate by comparing the Wesenheit magnitudes of classical Cepheids occupying metal-rich and metal-poor fields in M101.  Evaluation (1) is employed to assess the conclusion by \citet[][their Fig.~28]{ss10} that the slope of the Wesenheit function is sensitive to abundance changes.  \citet{mag09} analyzed the same HST images for M101\footnote{Taken \emph{verbatim} from their published AAS abstract, see also http://www.stsci.edu/observing/phase2-public/11297.pdf} and reached alternate conclusions, namely that there is no significant dependence on metallicity for the slope, and a comparatively small dependence on the zero-point of the P-L relation exists \citep[see also][]{fr01,st11b}. Published results for other galaxies \citep[e.g., NGC 5253,][]{gi00} are likewise interpreter/sample selection/pipeline dependent, thereby highlighting an often uncharacterized source of uncertainty. 

The dissenting (alternate) view conveyed here concerning a sizable $\gamma$-correction does not mitigate the broader significance of the \citet{ss10} and \citet{ge11} results. \citet{ss10} discovered $\sim 10^3$ classical Cepheids in M101, thereby exceeding existing records for the number of extragalactic Cepheids detected in a particular galaxy beyond the Local Group. Indeed, it is hoped that their approach may be applied to discover countless Cepheids in additional galaxies.  \citet{ge11} demonstrated the pertinence of the Large Binocular Telescope for fostering extragalactic Cepheid research.  Both studies present seminal results.

\begin{figure}[!t]
\epsscale{.8}
\plotone{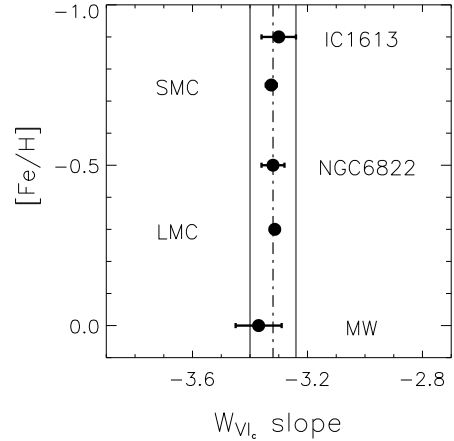}
\caption{\small{The slope of the {\it VI}$_c$ Wesenheit function as inferred from  ground-based photometry of classical Cepheids in the Milky Way (equation~\ref{eqn-mw2}), LMC, NGC 6822, SMC, and IC 1613.  The slope of the {\it VI}$_c$ Wesenheit function is insensitive to metallicity over the range examined.}}
\label{fig-ws}
\end{figure}

\section{{\rm \footnotesize EVIDENCE FOR THE ABSENCE OF A METALLICITY EFFECT IN {\it VI}$_C$}}
\label{s-m}
Firstly, evidence hitherto which indicate the relative insensitivity of $VI_c$ intrinsic color and Wesenheit functions to abundance changes are summarized, namely since such pertinent evidence is often overlooked. 

The results for evaluation (1) indicate that classical Cepheids in the Milky Way, LMC, NGC 6822, SMC, and IC 1613 follow a common {\it VI}$_C$ Wesenheit slope \citep[][see also \citealt{pi04,pg04,ma08,ma09b,so10}]{ma10b}, to within the uncertainties (Fig.~\ref{fig-ws}).  The galaxies span a sizable abundance baseline, thereby permitting a reliable determination of any trend.  Conversely, classical Cepheids in the Milky Way and SMC exhibit differing $BV$ Wesenheit slopes \citep[][see also \citealt{cc85}]{ma08,ma09b}.  The latter dependence appears tied to increased line blanketing in $BV$ \citep[][and references therein]{cc85}.  

The {\it VI}$_C$ Wesenheit results imply that the source for the following discrepancies is unrelated to a metallicity effect: the slope of the {\it VI}$_C$ Wesenheit function varies as a function of galactocentric distance for classical Cepheids in M101 and M106 \citep[e.g.,][their Fig.~28]{ss10}; the slope of the {\it VI}$_C$ Wesenheit function inferred from the \citet{sa04} Galactic calibrating sample\footnote{\citet{sa04} Galactic Cepheid calibration relied upon the best available data prior to the release of the HST parallaxes \citep{be07}.  Moreover, \citet{tu10} and \citet{st11a} subsequently revised the Galactic calibration, and continued revisions will invariably ensue.} is too steep, yielding distances for longer-period Cepheids which are artificially too large \citep[][see also \citealt{be07,vl07,st11a}]{ma10b}; and conversely, classical Cepheids in several SNe host galaxies exhibit too shallow a $VI_c$ Wesenheit slope and negative mean reddenings \citep[IC 4128, NGC 1309, NGC 3021,][and references therein]{ma10b}.

Regarding evaluation (2), the Wesenheit magnitude offset between RR Lyrae variables and classical Cepheids in the LMC, SMC, and IC 1613 agree to within the uncertainties \citep[][see also \citealt{ud01,pg04,pi04}]{ma10b}.  Distances and extinction estimates inferred for RR Lyrae, Type II Cepheid, and classical Cepheid variables in countless galaxies and globular clusters via metallicity-uncorrected period-based relations are comparable, and agree with results from DIRBE/IRAS dust maps \citep[e.g., M33 and M54,][]{ma09b,ma09d}.  

Concerning evaluation (3), results by several authors imply that the (spurious) brightening ($W_{VI_c}$) of extragalactic Cepheids as a function of decreasing galactocentric distance stems from the associated increase in photometric contamination, rather than as a result of increasing metal abundances \citep{mo00,ma01,mo04,bo08,ma09b,ma10b,br11}.  The surface brightness and stellar density increase near the central region, and hence the effects of crowding and blending cannot be ignored.  Further evidence presented below bolsters that assertion.  \citet{ss10} admit the impact of blending was not assessed in their analysis.  

\section{{\rm \footnotesize EVALUATING THE VIABILITY OF $\gamma\sim-0.8$ mag/dex}}
\label{s-m2}
The viability of a sizable metallicity correction ($\gamma\sim-0.8$ mag/dex) is now evaluated by applying it to distances inferred from classical Cepheids in the Magellanic Clouds via the Galactic calibration.  \citet{be07} cite a Galactic {\it VI}$_C$ Wesenheit function characterizing 10 nearby classical Cepheids with HST parallaxes as:
\begin{eqnarray}
W_{VI_c,0}=(-3.34\pm0.17) \log {P_0} -2.52 
\label{eqn-mw}
\end{eqnarray}
where $W_{VI_c,0}$ is the absolute Wesenheit magnitude and $\log{P_0}$ is the pulsation period tied to the fundamental mode. \citet{be02,be07} established HST parallaxes for the classical Cepheids RT Aur, T Vul, FF Aql, $\delta$ Cep, Y Sgr, X Sgr, W Sgr, $\beta$ Dor, $\zeta$ Gem, and $\ell$ Car. \citet{tu10} noted that the period-luminosity relation inferred from classical Cepheids in open clusters \citep[e.g., DL Cas/NGC 129,][]{tu92} matches that established from the \citet{be07} sample.  Moreover, the HST parallaxes were likewise corroborated by \citet{vl07} using revised Hipparcos parallaxes.  \citet{ma11} established precise $JHK_s$ ZAMS distances to 7 of 9 benchmark open clusters that agree with the revised Hipparcos estimates \citep{vl09}. In summary, the reliability of the HST parallaxes is supported by independent means.  

\citet{ma11b} supplemented the HST calibration with $21$ Galactic cluster Cepheids \citep{tu10} and obtained:
\begin{eqnarray}
W_{VI_c,0}=(-3.37\pm0.08) \log {P_0} -(2.48\pm0.08) 
\label{eqn-mw2}
\end{eqnarray}
The hybrid Galactic Wesenheit function includes the revised parameters for the classical Cepheid TW Nor in the open cluster Lyng{\aa} 6, which stemmed from an analysis of new VVV $JHK_s$ photometry for the cluster \citep{mi10,mb11,ma11b}.  That result agrees with the revised distance  established from the infrared surface brightness technique \citep{st11a}.  The short period classical Cepheid SU Cas was excluded from the derivation since its parameters are being revised by Turner \citep[see also discussion in][]{st11a}. 

{\it VI}$_C$ Wesenheit functions determined by \citet{so08,so10} that characterize $\ge10^3$ fundamental mode classical Cepheids in the Magellanic Clouds are:
\begin{eqnarray}
\nonumber
W_{VI_c}({\rm LMC})=(-3.314\pm0.009) \log {P_0} +(15.838\pm0.006)   \\
\nonumber
W_{VI_c}({\rm SMC})=(-3.326\pm0.019) \log {P_0} +(16.383\pm0.014) \\
\label{eqn-lmc}
\end{eqnarray}	
An analogous slope describes Galactic and Magellanic Cloud classical Cepheids which span $\rm{[Fe/H]}\sim 0 \rightarrow -0.33 \rightarrow -0.75$ (equations \ref{eqn-mw2}, \ref{eqn-lmc}, Fig.~\ref{fig-ws}). The coefficients and zero-points of the functions were confirmed by \citet{ma09b} and \citet{ng09}.  The distance modulus follows from subtracting the Wesenheit function inferred for a target population from the Galactic calibration:
\begin{eqnarray}
\nonumber
W_{VI_c,0}=W_{VI_c}-\mu_0 \\
\mu_0=W_{VI_c}-W_{VI_c,0}
\end{eqnarray}
Evaluating $\mu_0$ for the LMC and SMC by subtracting equation (\ref{eqn-lmc}) from equation (\ref{eqn-mw}) yields: $\mu_0\sim18.4$ and $\mu_0\sim18.9$, accordingly.  \citet{mot06} cite mean abundance estimates for classical Cepheids in the Magellanic Clouds as: $\rm{[Fe/H]}_{LMC}=-0.33\pm0.13$ and $\rm{[Fe/H]}_{SMC}=-0.75\pm0.08$ \citep[see also][]{lu98}.  The resulting distance modulus corrections owing to abundance differences between Galactic and Magellanic Cloud classical Cepheids are: $\Delta \mu_{0,\gamma}(LMC)\sim-0.26$ and $\Delta \mu_{0,\gamma}(SMC)\sim-0.60$, for $\gamma\sim-0.8$ mag/dex \citep{ss10}.  The metallicity corrected distance estimates for the LMC and SMC are therefore:  $\mu_{0,\gamma}\sim18.1$ and $\mu_{0,\gamma}\sim18.3$, respectively.  The results are  concerning since they imply that the Magellanic Clouds are $\sim20$\% nearer than inferred from a mean of $> 300$ published estimates, including the independent distance determined below from a Universal Wesenheit template.  Furthermore, the separation between the Clouds ($\Delta \mu_{0,\gamma}\sim0.2$) is approximately half the canonical estimate.

\begin{deluxetable}{cccccc}
\tablewidth{0pt}
\tabletypesize{\small}
\tablecaption{\small{Distances for the Magellanic Clouds}}
\tablehead{ & \colhead{$\rm{[Fe/H]}$\tablenotemark{1}} & \colhead{$\mu_0 (\gamma \sim 0$ mag/dex)\tablenotemark{2}} & \colhead{$\mu_0 ( \gamma \sim -0.8$  mag/dex)\tablenotemark{2}} & \colhead{NED-D\tablenotemark{3}} & \colhead{$\mu_{0,uwt}$\tablenotemark{4}} }
\startdata
LMC & $-0.33\pm0.13$ & $18.4$ & $18.1$ & $18.46\pm0.01(\sigma_{\bar{x}}) \pm 0.15 (\sigma )$ & $18.40\pm0.08(\sigma_{\bar{x}})$   \\
SMC & $-0.75\pm0.08$ & $18.9$ & $18.3$  & $18.86\pm0.02(\sigma_{\bar{x}}) \pm 0.18 (\sigma )$ & {\it --} \\
\enddata
\tablenotetext{1}{\scriptsize{Mean Cepheid abundances from \citet{mot06}, which agree with the earlier determinations by \citet{lu98}.}}
\tablenotetext{2}{\scriptsize{Metallicity ($\gamma$) corrected distances established from a {\it VI}$_C$ Galactic classical Cepheid Wesenheit function (equation \ref{eqn-mw}).}}
\tablenotetext{3}{\scriptsize{Distances for the Magellanic Clouds tabulated from $>3\times10^2$ published estimates (NED-D).  A mean LMC distance derived from additional published estimates is forthcoming (Steer et al., in prep.).}}
\tablenotetext{4}{\scriptsize{Inferred from a universal Wesenheit template featuring 11 nearby $\delta$ Scuti, SX Phe, RR Lyrae, and Type II Cepheids variables with HST/Hipparcos parallaxes.}}
\label{table1}
\end{deluxetable}

A consensus distance for the LMC may be derived using the NASA/IPAC Extragalactic Database NED-D \citep{ms07}.  That compilation features redshift-independent distances for $10^4$ galaxies.  NED-D contains $>3\times10^2$ distance estimates for the Magellanic Clouds, excluding those established from classical Cepheids \citep[e.g.,][]{st11b}.  The mean values for the LMC and SMC are: $\mu_0=18.46\pm0.01(\sigma_{\bar{x}}) \pm 0.15 (\sigma )$ and $\mu_0=18.86\pm0.02(\sigma_{\bar{x}}) \pm 0.18 (\sigma )$.  The results disagree with the metallcity corrected distances established using $\gamma\sim-0.8$ mag/dex (Table~\ref{table1}).  By contrast, the results inferred from NED-D agree with the $W_{VI}$-based distances uncorrected for metallicity differences between Magellanic Cloud and Galactic classical Cepheids (Table~\ref{table1}).

The distance to the LMC established via a universal Wesenheit template \citep{ma10c,ma11} featuring 11 nearby $\delta$ Scuti, SX Phe, RR Lyrae, and Type II Cepheids variables with HST/Hipparcos parallaxes is: $\mu_0=18.40\pm0.08(\sigma_{\bar{x}})$.  The prototypes RR Lyrae and SX Phe are included in the calibration owing to the availability of HST and revised Hipparcos parallaxes \citep{be02b,vle07}. The distance was inferred by matching the calibrated Wesenheit template to OGLE observations of LMC variables \citep[e.g.,][for $\delta$ Scuti stars]{po10}.  The impetus for the universal Wesenheit template is to employ the statistical weight of the entire variable star demographic to establish precise distances, constrain pulsation modes, and to provide broader context to identify peculiarities among certain variables \citep{ma10c,ma11}. The reddening-free nature of the Wesenheit approach obviates the propagation of uncertainties tied to tentative total/differential extinction corrections, ensuring that further calibration may ensue directly from published or forthcoming geometric-based distances (masers, HST, VLBA, GAIA).  The distance established to the LMC from the Wesenheit template matches that from NED-D and a calibration based on classical Cepheids only (equations~\ref{eqn-mw} \& 2).  

\section{{\rm \footnotesize CONCLUSION \& FUTURE RESEARCH}}
 A sizable metallicity correction ($\gamma\sim-0.8$ mag/dex) was evaluated by applying it to distances established for classical Cepheids in the Magellanic Clouds via the Galactic $W_{VI_c}$ function (equation~\ref{eqn-mw}, Table~\ref{table1}). The ensuing metallicity corrected distances for the Magellanic Clouds are in significant disagreement with estimates from countless indicators (Table~\ref{table1}).  In tandem with the evidence summarized in \S \ref{s-m}, the results indicate that variations in chemical composition among Cepheids are a comparatively insignificant source of uncertainty for $W_{VI}$-established distances and determinations of $H_0$.  Metallicity corrections seem unnecessary for $W_{VI_c}$-based distances, and consequently the observed apparent brightening ($W_{VI_c}$) of extragalactic Cepheids with decreasing galactocentric distance \citep{ke98,ss10} likely stems from the associated increase in surface brightness and stellar density toward the galaxy center \citep[][see also \citealt{mo00,ma01,mo04,bo08,br11}]{ma09b,ma10b}.  The disagreement with \citet{ss10} regarding the nature of the {\it VI}$_C$ Wesenheit function's metallicity dependence does not mitigate the significant accomplishments achieved in their comprehensive analysis of M101.  \citet{ss10} discovered $\sim10^3$ classical Cepheids using hybrid and modified variable search routines. Furthermore, the results for M101 provide additional empirical constraints on photometric contamination (blending/crowding), which in harmony with the challenges of establishing precise, standardized, multi-epoch, multi-band photometry: constitute a significant source of uncertainty for extragalactic Cepheid distances.  
 
Future research shall aim to assess the viability of establishing classical Cepheid calibrators from their membership in spiral arms, which are likewise delineated by young open clusters \citep[e.g.,][]{ma09,ma09c}.  A $\ell$-distance relation, where $\ell$ is the galactic longitude, can be inferred from open clusters with precise parameters derived via VVV/UKIDSS/2MASS $JHK_s$ photometry \citep{lu08,mi10}. The dispersion for the $\ell$-distance relation is particularly limited for certain sight-lines, such as toward the Carina arm \citep{ma09,ma09c,ma10}. The $\ell$-distance relation may be subsequently  applied to classical Cepheids which are spiral arm members: e.g., SZ Vel and RY Vel.  The aforementioned classical Cepheids exhibit pulsation periods of 14 and 28$^{\rm d}$ accordingly, and would bolster the longer period regime of the Galactic calibration which is comparatively undersampled \citep[see \citealt{be07} or Fig.~3 in][]{ma11b}.  Securing additional long-period calibrators would mitigate present uncertainties associated with the slope of the Galactic Wesenheit function (equation~\ref{eqn-mw2}), and permit a more reliable determination of the parameter's insensitivity to chemical composition (Fig.~\ref{fig-ws}).  Moreover, longer period classical Cepheids are particularly important since they are most often sampled in remote galaxies owing to their increased luminosity relative to shorter period Cepheids.  The Hubble flow dominates proper motions for remote galaxies, thereby minimizing uncertainties tied to the latter parameter and hence $H_0$. Indeed, the debate surrounding the SNe Ia scale and $H_0$ \citep{fr01,sa04} centres in part around the intrinsic parameters of longer period Cepheids \citep[e.g., see Fig.~3 in][]{ma10b}.  

The results presented here emphasize the importance of characterizing and correcting spurious (e.g., contaminated) photometry tied to distant Cepheids when determining the metallicity dependence on $W_{VI_c}$ and $H_0$. Admittedly, additional research on the topic is required.
 
\subsection*{{\rm \scriptsize ACKNOWLEDGEMENTS}}
\scriptsize{DM is grateful to the following individuals and consortia whose efforts and surveys were the foundation of the study: OGLE, the Araucaria project, F. Benedict, F. van Leeuwen, L. Berdnikov, I. Steer (NED-D), and staff at the CDS, arXiv, and NASA ADS.  WG gratefully acknowledges financial support for this work from the Chilean Center for Astrophysics FONDAP 15010003, and from the BASAL Centro de Astrofisica y Tecnologias Afines (CATA) PFB-06/2007. The VVV survey is conducted with the VISTA telescope under the ESO Public Survey programme ID 179.B-2002.}

\end{document}